\documentclass[12pt]{revtex4}
\usepackage{amsmath}
\usepackage{amssymb}
\usepackage{graphics}

\newcommand\ket[1]{\left|#1\right>}
\newcommand\bra[1]{\left<#1\right|}

\newcommand\Schr{Schr\"odinger\,}

\newcommand\dg{^\dagger}
\newcommand\pr{^\prime}

\newcommand\ro{\hat\rho}
\newcommand\Ho{\hat H}
\newcommand\xb{\mathbf{x}}
\newcommand\xbo{\hat{\xb}}
\newcommand\pb{\mathbf{p}}
\newcommand\pbo{\hat{\pb}}
\newcommand\pbi{\pb_i}
\newcommand\pbf{\pb_f}
\newcommand\kb{\mathbf{k}}
\newcommand\kbi{\kb_i}
\newcommand\kbf{\kb_f}
\newcommand\Ecal{^\mathcal{E}}

\begin{document}
\title{Decoherence and the puzzle of quantum Brownian motion in a gas}   
\author{Lajos Di\'osi}
\affiliation{Wigner Research Centre for Physics\\
H-1525 Budapest 114. P.O.Box 49, Hungary}
\date{\today}
\begin{abstract}
Heinz-Dieter Zeh's discovery that  the motion of macroscopic objects can not,
under typical conditions, follow the \Schr~equation necessitates a suitably
modified dynamics. This unfolded a long-lasting puzzle in the open quantum system 
context: what is the quantum counterpart of the classical Brownian motion in a gas.
Presented is a criticism and an open-end discussion of the quantum linear Boltzmann 
and quantum Fokker-Planck equations --- with constant respect for foundational research.
\end{abstract}

\maketitle
\section{Introduction}
The long time of ignorance after Zeh's publications {\cite{Zeh70,Zeh71} in 1970-71 on environmental
decoherence got broken by Wigner. 
He summarized his own revelation and Zeh's iscovery \cite{Wig83} as follows:
\emph{This writer's earlier belief that the physical apparatus' role can
always be described by quantum mechanics [\dots] implied that the
"collapse of the wave function" takes place only when the observation 
is made by a living being - a being clearly outside of the scope of
our quantum mechanics. The arguments which convinced me that
quantum mechanics' validity has narrower limitation, that it is
not applicable to the description of the detailed behavior of
macroscopic bodies, is due to D. Zeh. (1971) [\dots]. The point is that
a macroscopic body's inner structure, i.e. its wave function, is
influenced by its environment in a rather shot time even if it is
in intergalactic space. Hence it can not be an isolated system [\dots]. }   
Wigner raised the question: \emph{Can an equation for the time-change of 
the state of the apparently not-isolated system be proposed?}

Nowadays, after decades, the answer is part of the theory
of open quantum systems. But in 1983, it was a novelty that Wigner
showed a master equation for the massive object's density matrix, modeling
the decoherence of its rotational motion:
\begin{equation}\label{Weq}
\frac{d\ro}{dt}=-\frac{i}{\hbar}[{\Ho},\ro
               -\sum_{\ell m}\varepsilon_{\ell}
                     \left[\hat L_{\ell m},[\hat L_{\ell m},\ro]\right]~,
\end{equation}
where ${\Ho}$ is the Hamiltonian of the macroscopic object and $\hat L_{\ell m}$ are
the multipole operators of its angular momentum, 
strengths of their decoherence are given by the parameters $\varepsilon_{\ell}$. 

In 1985, Joos and Zeh (JZ) found \cite{JooZeh85} that decoherence of the center-of-mass position $\xbo$
would be more typical and, what is important,  its derivation is simple.

\section{Equation of positional decoherence}
Consider a macroscopic object, e.g. a dust, of mass $M$ under the influence of incoming plane
waves of particles, e.g.  molecules,  of mass $m\ll M$ that are scattered independently by the 
macroscopic object. 
JZ took the following unitary transition per single collisions, valid if $M\rightarrow\infty$:
\begin{equation}\label{JZcoll}
\ket{\pbi}\otimes\ket{\kbi}\Rightarrow
\ket{\pbi}\otimes\ket{\kbi}+\frac{i}{2\pi k_i}\int d\kbf
              f(\kbf,\kbi)\delta(k_f-k_i)\ket{\pbf}\otimes\ket{\kbf},
\end{equation}
$\pb_{i/f},\kb_{i/f}$ are the initial/final momenta of the object and the particle, respectively,
where $\pbf=\pbi+\kbi-\kbf$ ensures momentum conservation, $f$ is the standard scattering amplitude. 
The authors pointed out that
repeated scatterings of the incoming particles contribute to gradual
localization of the object, i.e., the off-diagonal terms of the positional density matrix 
$\rho(\xb\pr,\xb)$ become damped. If the distribution $\rho^{\mathcal{E}}(\kb_i)$ of the 
incoming environmental particles is isotropic then the collisions contribute to the
following master equation:
\begin{equation}\label{JZeq}
\frac{d\ro}{dt}=-\frac{i}{\hbar}[{\Ho},\ro]-\Lambda[\xbo,[\xbo,\rho]]
\end{equation}
valid if the coherent extension of the object's position is much smaller than the
wavelength of the particles:
 \begin{equation}\label{smallDx}
 \vert\xb\pr-\xb\vert\ll\hbar/k.
\end{equation}
JZ determined the parameter $\Lambda$ of localization rate:
\begin{equation}\label{Lambda}
\Lambda=\frac{1}{\hbar^2}\times\mbox{incoming flux of particles}\times k^2 \sigma_{eff}.
\end{equation}
They calculated the effective cross section $\sigma_{eff}$ from the
differential cross section $\vert f\vert^2$.

The JZ master equation (\ref{JZeq}) is paradigmatic in decoherence theory. It describes the
gradual damping the off-diagonal elements of the positional density
matrix $\rho(\xb\pr,\xb)$  which is called positional decoherence on one hand
and yields localization of the coherent extension of the object on the other. 
In 1990 Gallis and Fleming \cite{GalFle90} revisited the considerations of JZ 
and refined their derivation of positional decoherence and its rate $\Lambda$.  
It is not clear when, lately, was the localization rate $\Lambda$ 
related to the classical diffusion coefficient for the first time. But the research
moved to that direction.

\section{Quantum Brownian motion in a gas}
The JZ master equation (\ref{JZeq}) has an alternative
interpretation, independent of and older than the concept of decoherence.
It corresponds to momentum diffusion of the Brownian object, with
the coefficient $D_p$ of momentum diffusion:
\begin{equation}
\Lambda=\frac{D_p}{\hbar^2}~.
\end{equation}
As a beneficial consequence, to obtain and understand the dynamics of decoherence,
also to complete the JZ master equation (\ref{JZeq}) by a term of friction,
we could have used the standard quantum theory of Brownian motion in a gas. 
Just this standard theory did not exist at the time. And it has since remained problematic
despite efforts of a community of researchers including myself. 
We all were motivated by our foundational interest in the quantum behavior of 
macroscopic objects under the influence of their uncontrollable environments. 
The efforts \cite{Dio95,Vac01,Hor06,HorVac08,VacHor09} started with 1995
and culminated in the Vacchini-Hornberger review \cite{VacHor09}. These autors say
\emph{... the seminal
paper on decoherence by Joos and Zeh [...], seeking to explain the absence of quantum delocalization in a dust particle by
the scattering of photons and air molecules, derived and studied what the authors called a Boltzmann-type master equation.
Two decades later, the long quest for the characterization of the phenomenon of collisional decoherence has now reached a
mature theoretical description, permitting its quantitative experimental confirmation}.
Let me outline the story, in my  ---selective and certainly subjective--- interpretation. 

In 1995 \cite{Dio95}, without mentioning  my foundational motivations, I asked the question: what is
the quantum Brownian dynamics of  the dust  in a dilute gas at thermal equilibrum? First I solved 
the classical problem by the linear variant of the classical Boltzmann-equation where  the 
molecule-molecule collision term is just replaced by the dust-molecule collision term. 
Unlike the classical case, the derivation of the quantum linear Boltzmann equation (QLBE)
was not straightforward.
Quantum mechanically, a single  collision corresponds to the following unitary transition,
generalizing (\ref{JZcoll}) for finite $M$:
\begin{equation}\label{Dcoll}
\ket{\pbi}\otimes\ket{\kbi}\Rightarrow
\ket{\pbi}\otimes\ket{\kbi}+\frac{i}{2\pi k_i^*}\int d\kbf^*
              f(\kbf^*,\kbi^*)\delta(k_f^*-k_i^*)\ket{\pbf}\otimes\ket{\kbf}~,
\end{equation}
$\kb_{i/f}^*$ are the initial/final momenta of the particle, respectively, in the
center-of-mass frame. As before,  $\pbf=\pbi+\kbi-\kbf$ ensures momentum conservation,
total energy conservation is ensured by the delta-function. When imposing the distribution 
$\rho\Ecal(\kb_i)$ of the gas molecule momenta, 
I had to introduce a heuristic \emph{maneouvre of square-root} (MSqR) otherwise the correct 
mathematical structure \cite{Goretal76,Lin76}
of the desired quantum master equation wouldn't have been achieved. The MSqR was
equivalent to a deliberate adding off-diagonal elements to the standard diagonal density 
matrix $\rho\Ecal(\kb,\kb\pr)\propto\rho\Ecal(\kb)\delta(\kb-\kb\pr)$ of the ideal
gas molecules. The choice was
\begin{equation}\label{MSqR}
\rho\Ecal(\kb,\kb\pr)=\sqrt{\rho\Ecal(\kb)}\sqrt{\rho\Ecal(\kb\pr)}~.
\end{equation}
This MSqR and some other simple assumptions led to the first QLBE of Brownian
motion in a gas. In the diffusion limit it yields the quantum Fokker-Planck equation (QFPE):
\begin{equation}\label{QFP}
\frac{d\ro}{dt}=-\frac{i}{\hbar}[{\Ho},\ro]
                              -\frac{D_p}{\hbar^2}[\xbo,[\xbo,\ro]]
                              -i\frac{\eta}{2\hbar}[\xbo,\{\pbo,\ro\}]
                              -\frac{D_x}{\hbar^2}[\pbo,[\pbo,\ro]]~.
\end{equation}
The coefficients of momentum diffusion $D_p=\eta Mk_BT$ and friction $\eta$ correspond to those in
the classical Fokker-Planck equation:
\begin{equation}\label{FPE}
 \frac{d\rho}{dt}=\{H,\rho\}_{Poisson}
                              +D_p\left(\frac{\partial}{\partial\pb}\right)^2\rho
                              +\eta\frac{\partial}{\partial\pb}\pb\rho~.
\end{equation}

However, the quantum version (\ref{QFP}) contains a strange term of \emph{position diffusion}
which would be nonsense classically. Position diffusion of the Brownian object is a pure
quantum effect,  the celebrated GKLS theorem \cite{Goretal76,Lin76} puts the following
lower bound on the coefficient of position diffusion:
\begin{equation}\label{Lindblad}
\frac{D_x}{\hbar^2}\geq\frac{\eta^2}{4D_p}=\frac{\eta}{4Mk_BT}~,
\end{equation}
which was satisfied in \cite{Dio95} by construction. Hornberger \cite{Hor06}, applying the MSqR,
found an ambiguity ---nicely elucidated later in \cite{VacHor09}--- 
and derived an alternative QLBE.  His was more natural than mine, in particular because
his QLBE  had the minimum possible value of $D_x$ [cf. (\ref{Lindblad})], i.e., the minimum rate 
of the strange position diffusion.

The context of quantum Brownian motion theory, the related results achieved by physicists
mostly working on quantum foundations otherwise, were summarized in 2009 \cite{VacHor09}
by Vacchini and Hornberger.
The QLBE of Hornberger \cite{Hor06} seemed to be the true and ultimate quantum version of
the classical linear Boltzmann equation.  But soon, an elementary argument  of decoherence 
popped up and questioned it together with all previous versions, including mine.

\section{Complete momentum decoherence (CMD)}
To understand the overlooked dramatic phenomenon indicated by the title above, we only need
the momentum and energy conservation of collision in one dimension first:
\begin{eqnarray}
p_i+k_i&=&p_f+k_f~,\label{pcons}\\
\frac{p_i^2}{2M}+\frac{k_i}{2m}&=&\frac{p_f^2}{2M}+\frac{k_f}{2m}~.\label{econs}
\end{eqnarray}
We express the final momentum of the the mass $M$ in the following form:
\begin{equation}\label{pfkikf}
p_f=\mu_+ k_i+\mu_- k_f~,
\end{equation}
where $\mu_\pm=(M/m\pm1)/2$.
Observe the initial momentum $p_i$ of the mass $M$ canceled!
Its final momentum depends on the initial $k_i$ and final  $k_f$ of the scattered mass $m$!
This fact yields  a crisis quantum mechanically. Observe that the reduced post-collision state
of the mass $M$ remains the same if we measure the post-collision momentum $k_f$ of the other mass.
Assume we do so and measure $k_f$.
Then the above expression of $p_f$ means that we  measure the final momentum $p_f$
of the mass $M$ as well and, as a consequence, momentum superpositions for the mass 
$M$ can exist no more after a single collision! \emph{Any single collision causes complete momentum
decoherence}  of the mass $M$. 

This trivial fact of CMD surfaced in 2009 \cite{Dio09} and  in 2010 \cite{KamCre10}) in the general case of 
the three-dimensional collision (\ref{Dcoll})  where the expression (\ref{pfkikf}) survives for the 
components of $\pb_f,\kb_i,\kb_f$ parallel to the momentum transfer $\kb_f-\kb_i$ only:
\begin{equation}\label{pfkikfparallel}
p_f^\parallel=\mu_+ k_i^\parallel+\mu_- k_f^\parallel~.
\end{equation}
CMD  in all three components of $\pb_f$  requests just three collisions in a row.
CMD is obviously unphysical. It would, in particular, 
suggest a divergent coefficient $D_x=\infty$ of position diffusion in the QFPE (\ref{QFP}).

The obligate question follows: how did the derivations of QLBEs from 1995 over fifty years 
got finite $D_x$  against the trivial CMD which imposes $D_x=\infty$. How did they
regularize the divergent position diffusion? 

\section{Collision and methods revisited}
We go back to the type of elementary considerations of JZ, this time taking
the exact collision kinematics like (\ref{Dcoll}) into the account, instead of the approximate (\ref{JZcoll}).     
To detect CMD and the role of the MSqR in its regularization, it is sufficient if simplify the derivations 
from three to one dimension. 
First, let us find the
counterpart of (\ref{Dcoll}) in one dimension. Assume, again for simplicity,  the repulsive hard-wall potential 
between the dust and a molecule so that we can ignore that they tunnel through each.
Then the unitary transition (\ref{Dcoll}) in a collision reduces to:
\begin{equation}\label{oneDcoll}
\ket{p}\otimes\ket{k}\Rightarrow\ket{p+2k^*}\otimes\ket{k-2k^*}~.
\end{equation}
[For brevity, we stop indicating that all momenta are the initial ones.]
Remember the center-of-mass initial momentum $k^*=(Mk-mp)/(M+m)$. 
If the initial state of the dust is a superposition of momentum eigenstates, the
transition of an off-diagonal element of the density matrix reads:
\begin{equation}
\ket{p}\bra{p\pr}\otimes\ket{k}\bra{k}\Rightarrow
\ket{p+2k^*}\bra{p\pr+2k^*}\otimes\ket{k-2k^*}\bra{k-2k^{*\prime}}~,
\end{equation}
where $k^{*\prime}=(Mk-mp\pr)(M+m)$. Take the partial trace of both sides, yielding
\begin{equation}
\ket{p}\bra{p\pr}\Rightarrow0~,
\end{equation}
because the two post-collision states of the molecule, scattered on $\ket{p}$ and $\ket{p\pr}$ resp., 
became orthogonal:   
\begin{equation}\label{orthog}
\langle{k_i-2k^{*\prime}}\ket{k-2k_i^*}=0~.
\end{equation}
This proves CMD analytically and confirms the previous measurement theoretical argument:
\emph{Any single collision causes CMD} of the dust.  This can not 
happen in the reality since it would completely delocalize the wave function.
It is now obvious that the said two post-collision states of the molecule should overlap! 

Our derivations \cite{Dio95,Hor06,HorVac08,VacHor09} of QLBE's created this overlap 
formally via the MSqR (\ref{MSqR}), without any awareness or reference to the above 
physical background.
We assumed an environmental ideal gas, i.e., a mixture of plane waves of thermal distribution
$\rho\Ecal(k)\propto\exp(-k^2/mk_BT)$ of temperature $T$.
But at a certain later stage towards the QLBE, we took the MSqR (\ref{MSqR}) and
postulated the following density matrix:
\begin{equation}\label{rhokkpr}
\rho\Ecal(k,k\pr)=\sqrt{\rho\Ecal(k)}\sqrt{\rho\Ecal(k\pr)}~,
\end{equation}
which represents a single normalized central real Gaussian wave function
$\psi\Ecal_k\propto\exp(-k^2/2mk_BT)$, i.e., a central real 
Gaussian wave packet standing at the origin:
\begin{equation}\label{psix}
\psi\Ecal(x)\propto\exp\left(-\frac{mk_BTx^2}{2\hbar^2}\right)~.
\end{equation} 
This single pure state served as an effective \emph{substitute} of the single molecule mixed state in the ideal gas.
The translation invariance was lost obviously. Nonetheless it became restored since all
plane wave components$\ket{k}$ were assumed to collide with the dust: 
\begin{equation}\label{oneDcollpsi}
\int dk\psi\Ecal_k\ket{p}\otimes\ket{k}\Rightarrow\int\psi\ket{p+2k^*}\otimes\ket{k-2k^*}~.
\end{equation}
It is of course hard to interpret this assumption but it was implicit in all derivations
and, most importantly, restored the translation invariance of the resulting QLBE.

Due to the MSqR (\ref{rhokkpr}) the two post-collision states of the molecule are, unlike in (\ref{orthog}),
no more orthogonal,
they overlap, the effect of CMD disappears,  gives its role to finite momentum decoherence .  Detailed calculations, omitted here, yield the 
QFPE (\ref{QFP}) with the the standard momentum diffusion $D_p=\eta Mk_BT$ and the
finite coefficient of position diffusion (momentum decoherece) $D_x$ saturating the
constraint (\ref{Lindblad}).  In  all historic  QLBEs \cite{Dio95,Hor06,HorVac08,VacHor09}
it is the MSqR that removes the divergence of $D_x$. 

Are these $D_x$'s physical? In view of  the meaning of the MSqR that a
standing wave packet substitutes the ideal gas single-molecule density matrix,
a finite $D_x$ may well be an artifact of the MSqR, as suggested by \cite{Dio09,KamCre10}.

\section{Farwell MSqR}
What other,  more physical mechanism could explain the finite physical momentum decoherence (position
diffusion) if it is tractable at all via the independent collisions. Should one improve on the single molecule
density matrix  of the ideal gas by taking molecule-molecule interactions into the account? Unfortunately,
one should not. The diagonal form $\rho(\pb,\pb\pr)\propto\delta(\pb-\pb\pr)$ remains because the
due translation invariance of the gas equilibrium state. To mitigate CMD,  playing with the quantum state 
$\rho\Ecal(\kb,\kb\pr)\propto\rho\Ecal(\kb)\delta(\kb-\kb\pr)$ of the particles is useless. We play with
the collision.

When MSqR turned out to be kind of unphysical elimination of CMD, the following consideration arose.
CMD assumes  idealized quantum scatterings that means, e.g., infinite intercollision time $\tau=\infty$.
If one takes the finite $\tau$ valid even in dilute gas then energy  conservation (\ref{econs}) in 
single collision becomes unsharp and CMD becomes relaxed. This was certainly a more justified
mechanism to mitigate CMD than the MSqR had been, I thought in \cite{Dio09}, and got
a finite coefficient $D_x$ of momentum decoherence (position diffusion):
\begin{equation}
D_x=\frac{1}{3}\left(\frac{\tau^2}{M}\right)D_p.
\end{equation} 
Hornberger and Vacchini \cite{HorVac10} claimed that the CMD issue was nonexistent in their 
QLBE \cite{Hor06,HorVac08,VacHor09} which contains the ultimate physics of quantum Brownian motion
---I disagreed \cite{Dio10}---   as long as  binary indepedent collisions are considered between the dust 
and the molecules. Also Kamleitner and Cresser \cite{KamCre10} blamed the idealization
of the scattering process for CMD and introduced a nonzero collision (interaction) time instead
of the idealized zero.  Apparently, no consensus has since been achieved as to the value of
$D_x$ neither to  the very existence of momentum decoheretnce (position diffusion).

This issue is not yet too burning since the effect is not testable currently. The experimental
significance of a non-zero $D_x$ was anticipated long time ago \cite{Jacetal99}, a possible test was
mentioned tangentially \cite{Beretal06}, a fundamental experiment \cite{Horetal03} used and 
confirmed the QLBE prediction for momentum diffusion only.

\section{Epilogue}
Many times, questioning the conservative and confirmed wisdom respecting quantum
mechanics turns out to be  unproductive.  Zeh's criticism was different
and changed our abstraction and practice about coherence in quantum theory.
I only wished to illustrate how Zeh's work, apart from its impact on foundations,
opened the Pandora's box of a standard unsolved problem independent of
foundations. What is our theory of a quantum Brownian particle in a gas? 
Theory ran into a puzzle that ---I'm afraid--- has remained unsolved so far.

\end{document}